\documentclass[12pt,preprint]{aastex}
\usepackage{graphicx, amsmath, amssymb, amsthm}
\usepackage{natbib}
\usepackage{apjfonts}
\usepackage{mathptmx}

\def\swift{{\it Swift}}

\begin{document}

\title{In search of progenitors for supernova-less GRBs 060505 and 060614: re-examination of their afterglows
\thanks{Based on observations collected at the European Organisation for Astronomical
Research in the Southern Hemisphere, Chile, under programs
077.D-0661 and 177.A-0591.}
}

\author{D.~Xu\altaffilmark{2,3}, R.~L.~C.~Starling\altaffilmark{3}, J.~P.~U.~Fynbo\altaffilmark{2},
J.~Sollerman\altaffilmark{2,4}, S.~Yost\altaffilmark{5}, D.~Watson\altaffilmark{2},
S.~Foley\altaffilmark{6}, P.~T.~O'Brien\altaffilmark{3}, J.~Hjorth\altaffilmark{2}}

\altaffiltext{2}{Dark Cosmology Centre, Niels Bohr Institute, University of Copenhagen,
Juliane Maries Vej 30, DK-2100 Copenhagen \O, Denmark; dong@dark-comology.dk}

\altaffiltext{3}{Department of Physics and Astronomy, University of Leicester, University
Road, Leicester LE1 7RH, UK}

\altaffiltext{4}{Department of Astronomy, Stockholm University, AlbaNova, 10691,
Stockholm, Sweden}

\altaffiltext{5}{Department of Physics, College of Saint Benedict, Saint John's
University, 37 South College Avenue, St. Joseph, Minnesota 56374, USA}

\altaffiltext{6}{UCD School of Physics, University College Dublin, Dublin 4, Ireland}

\begin{abstract}
GRB\,060505 and GRB\,060614 are nearby long-duration gamma-ray bursts (LGRBs) without
accompanying supernovae (SNe) down to very strict limits. They thereby challenge the
conventional LGRB-SN connection and naturally give rise to the question: are there other
peculiar features in their afterglows which would help shed light on their progenitors?
To answer this question, we combine new observational data with published data and
investigate the multi-band temporal and spectral properties of the two afterglows. We
find that both afterglows can be well interpreted within the framework of the jetted
standard external shock wave model, and that the afterglow parameters for both bursts
fall well within the range observed for other LGRBs. Hence, from the properties of the
afterglows there is nothing to suggest that these bursts should have another progenitor
than other LGRBs. Recently, {\it Swift}-discovered GRB\,080503 also has the spike $+$
tail structure during its prompt $\gamma$-ray emission seemingly similar to GRB\,060614.
We analyse the prompt emission of this burst and find that this GRB is actually a
hard-spike $+$ hard-tail burst with a spectral lag of 0.8$\pm$0.4 s during its tail
emission. Thus, the properties of the prompt emission of GRB\,060614 and GRB\,080503 are
clearly different, motivating further thinking of GRB classification. Finally we note
that, whereas the progenitor of the two SN-less bursts remains uncertain, \emph{the
core-collapse origin for the SN-less bursts would be quite certain if a wind-like
environment can be observationally established, e.g, from an optical decay faster than
the X-ray decay in the afterglow's slow cooling phase}.

\end{abstract}


%%%%%%%%%%%%%%%%%%%%%%%%%%%%%%%%%%%%%%%%%%%%%%%%%%%%%%%%%%%%%%%%%%%%%%%%%%%%%%%%%%%%%%
\section{Introduction}
Gamma-ray bursts (GRBs) are the most luminous explosions in the Universe since
the Big Bang. They fall into two (partially overlapping) populations according
to their observed duration: $\gamma$-ray durations (measured as the time in
which 90\% of the fluence is emitted) longer than 2~s are defined as long GRBs
(LGRBs) while bursts with duration shorter than 2~s are defined as short GRBs
(SGRBs; \citealt{Kouveliotou93}). It is widely accepted that at least the
majority of LGRBs are driven by the collapse of massive stars (e.g.,
\citealt{WB06}), although some LGRBs may be generated by the merger of compact
objects (e.g., \citealt{KR98,RRD03}). The strongest evidence for the collapsar
scenario is the detection of bright Ic SN component photometrically and
spectroscopically associated with nearby LGRBs such as GRB\,980425,
GRB\,030329, GRB\,031203, and XRF\,060218
(\citealt{Galama98,Hjorth03,Malesani04,Sollerman06,Pian06,Campana06}). On the
other hand, SGRBs may be powered by the merger of binary compact objects (e.g.
\citealt{Eich89,Narayan92}). This connection is observationally bolstered by
the association of some SGRBs with old stellar populations and lack of
accompanying bright SN components in cases such as GRB\,050509B
(\citealt{Hjorth05a}), GRB\,050709 (\citealt{Fox05,Hjorth05b}) and GRB\,050724
(\citealt{Berger05,Malesani07}). However, challenging this simple picture some
SGRBs displayed violent X-ray flares occurring at least $\sim100$ s after the
triggers, e.g., GRB\,050709 (\citealt{Fox05}) and GRB\,050724
(\citealt{Bart05,Campana06,Malesani07}). This suggests long-lasting activity of the
central engine and hence the current understanding of the GRB progenitor
mechanism may be too simple (e.g., \citealt{Fan05,Dai06,Ross07}). There is also
evidence for activity of the inner engine on much longer time-scales (several
days) for GRB\,050709 (\citealt{Watson06}) and GRB\,070707
(\citealt{Piranomonte08}).

The whole picture became more complicated after the discovery of GRB\,060505 and
GRB\,060614, because both bursts are nearby LGRBs according to the conventional taxonomy
but they are observationally not associated with SNe down to very strict limits
(\citealt{Fynb06,DellaValle06,Gal-Yam06}). In this sense, they share the expected
observational properties of both conventional LGRBs and SGRBs.

GRB\,060505 had a fluence of $(6.2\pm1.1)\times10^{-7}$~erg~cm$^{-2}$ in the $15-150$~keV
band, and a $T_{90}$ duration of $4\pm1$~s (\citealt{Hullinger06,McBreen08}). It was
found to be associated with a bright, star-forming \ion{H}{2} region within its host
galaxy at $z=0.089$ (\citealt{Ofek07,Thoene08}). In the compact-star merger scenario the
diameter of the \ion{H}{2} region and the location of the GRB within it suggest that the
delay time from birth to explosion of GRB\,060505 was $\lesssim 10$ Myr. This is
marginally matching the lower limit of the delay-time region for SGRBs
(\citealt{Ofek07}). On the other hand, the age of the \ion{H}{2} region of $\sim 6$ Myr
(\citealt{Thoene08}) is consistent with the expectation for core-collapse in a massive
star. The prompt emission of GRB\,060614 consisted of a hard-spectrum component lasting
$\sim5$~s followed by a soft-spectrum component lasting $\sim100$~s. \cite{Mangano07}
reported a photon index $\Gamma=1.63\pm 0.07$ for the time interval [-2.83,-5.62] s since
the BAT trigger ($\chi^2/{\rm dof}=48.2/56$) and $\Gamma=2.21\pm0.04$ for 5.62-97.0 s
since the BAT trigger ($\chi^2/{\rm dof}=40.9/56$). The fluences in the two components
are $(3.3\pm0.1)\times10^{-6}$~erg~cm$^{-2}$ and
$(1.69\pm0.02)\times10^{-5}$~erg~cm$^{-2}$ in the 15-350 keV band, respectively
(\citealt{Gehrels06}). Its host galaxy has a redshift of $z=0.125$ (\citealt{PBF06}). The
host of GRB\,060614 is very faint with an absolute magnitude of about $M_B=-15.3$
(\citealt{Fynb06}). Its specific star-formation rate is quite low, but within the range
covered by LGRB hosts (similar, e.g., to that of the GRB\,050824 host galaxy,
\citealt{Sollerman07}).

The spectral lag has been invoked as a quantity that can be used to classify bursts such
that SGRBs have zero lag and LGRBs fall on a well defined lag-luminosity relation
\citep{Norris00,Norris06}. For GRB\,060614 \cite{Gehrels06} found that the spectral lags
for the hard and soft components in the prompt $\gamma$-ray emission are both consistent
with zero lag, falling entirely within the range for SGRBs. For GRB\,060505, on the other
hand, \cite{McBreen08} found using the {\it Suzaku}/WAM and {\it Swift}/BAT data that the
spectral lag for the prompt $\gamma$-ray emission is $0.36\pm0.05$~s, consistent with a
LGRB identity. Furthermore, lags of LGRBs and SGRBs, regardless of their physical
origins, appear to overlap quite significantly according to statistics of 265 {\it Swift}
bursts (see Fig. 1 in \citealt{Bloom08}). Alternatively, the so-called Amati relation can
be used to provide hints on which class GRB\,060505 and GRB\,060614 belong to. According
to this relation derived from the observed GRBs with sufficient data, all SGRBs are
outliers because of their relatively higher $\nu F_\nu$ peak energy, while all LGRBs,
except the peculiar long GRB\,980425, are consistent with this relation. Amati et al.\
(2007) find that GRB\,060614 follows the relation whereas GRB\,060505 does not. Hence,
based on properties of the prompt emission other than the duration it seems impossible to
establish clear evidence about which class of bursts GRB\,060505 and GRB\,060614 belong
to.

To gain further insight on this topic, in this work we add our own observational data to
already published data and study the afterglows of GRB\,060505 and GRB\,060614. We aim to
determine from the afterglow properties whether these two bursts differ from other LGRBs,
besides being SN-less, to provide further clues to the nature of their progenitors. The
observations of the two afterglows and data reduction are presented in \S~2.1 and \S~2.2.
At the beginning of \S~3, we briefly introduce the leading external shock wave model
employed to explain GRB afterglows, and apply it to GRB\,060505 in \S~3.1. GRB\,060614
has been studied by \cite{Mangano07}; in \S~3.2 and \S~3.3 we re-analyze this burst and
provide analytical and numerical constraints on the afterglow parameters, respectively.
In \S~4 we discuss possible progenitors of these two bursts in comparison with the
recently discovered GRB\,080503, which we define as the first hard-spike $+$ hard-tail
{\it Swift} GRB.

Throughout this paper we use the notation $F_{\nu}(\nu,t) \propto t^{-\alpha}
\nu^{-\beta}$ for the afterglow monochromatic flux as a function of time, where $\nu$
represents the observed frequency and $\beta$ is the energy index which is related to the
photon index $\Gamma$ in the form of $\beta = \Gamma-1$. The convention $Q_x=Q/10^x$ has
been adopted in cgs. In addition, we consider a standard cosmology model with $\rm{H_0} =
70$ km s$^{-1}$ Mpc$^{-1}$, $\Omega_{\rm M} = 0.3$, and $\Omega_\Lambda = 0.7$.

\section{Observations and data reduction}
\subsection{GRB\,060505}
The field of GRB\,060505 was observed with the European Southern Observatory (ESO) Very
Large Telescope (VLT) and the FORS1 instrument on two epochs (see also \citealt{Fynb06}).
On May 6.4, slightly more than one day after the burst, the field was observed in the $B,
V, R, I$ and $z$ bands. In order to be able to subtract the underlying host galaxy, in
particular the hosting star-forming region within the host galaxy (\citealt{Thoene08}),
the field was observed again on September 14.2, again in the $B, V, R, I$ and $z$ bands.
The journal of observations is given in Table 1. Due to strong fringing and lack of
calibration data we decided not to include the $z$ band data in the analysis.

\begin{figure*}
\centerline{\includegraphics[width=16cm]{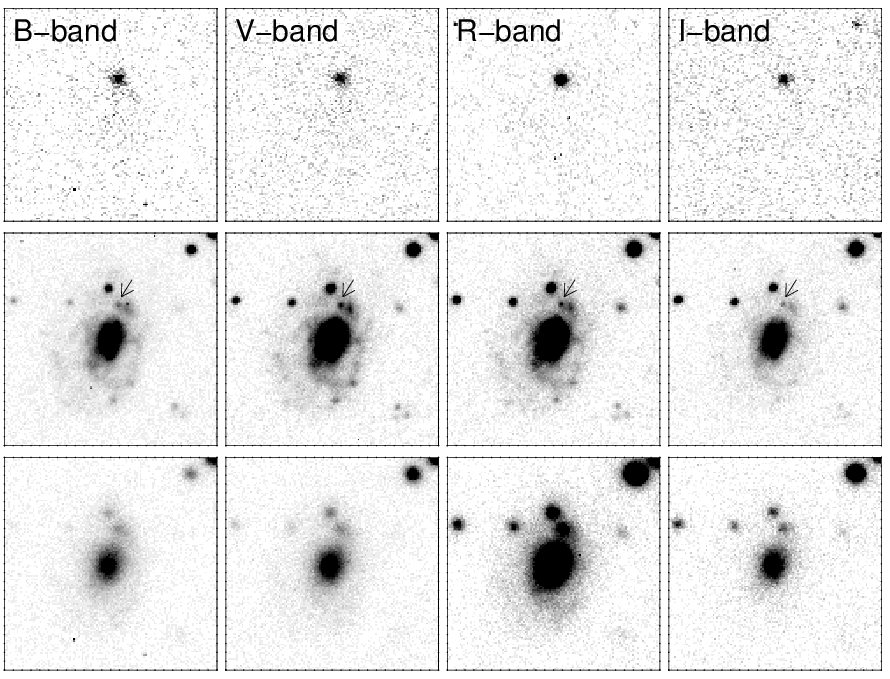}} \caption{ Multicolor imaging
of the afterglow and host galaxy of GRB\,060505. The top row shows the result of image
subtraction, that is, imaging on May 6.4 minus imaging on September 14. The middle row
shows the deeper, and better seeing imaging obtained on September 14, more than 4 months
after the burst, while the bottom row shows the imaging of the field on May 6.4, $\sim
1.125$ days after the burst. \label{Image_Subtraction}}
\end{figure*}

The optical data were corrected for bias and flat-fielded using standard
techniques. In order to subtract the underlying emission from the host galaxy
we used the ISIS software (\citealt{Alard98}). In Fig.~\ref{Image_Subtraction}
we show the result of the image subtraction. As can be seen, the afterglow is
clearly detected in all four bands. We then performed photometry on the
afterglow in the following way. We first duplicated an isolated, non-saturated
star in each of the the first epoch images to a new empty position such that it
also appeared in the subtracted image. We then used Daophot
(\citealt{Stetson87}) to perform relative photometry between the afterglow and
the star.  Finally, we obtained the photometry on the standard system by
measuring the magnitude of the comparison star using aperture photometry and
the photometric zeropoints obtained based on Landolt stars by the ESO
observatory calibration plan on the same night.

The \emph{Swift}/XRT data were processed in a standard way using the {\it
Swift} software version 29 (released 2008 June 29 as part of HEAsoft 6.5.1). We
also included in the analysis the \emph{Swift}/UVOT data points/upper limits
and the late X-ray data point using the ACIS-S detector on board the {\it
Chandra} X-ray observatory in \cite{Ofek07}.

\subsection{GRB\,060614} Table~\ref{0614Optical} shows the comprehensive
$R$-band data of the afterglow of GRB\,060614 from the Watcher 0.4m telescope, DFOSC at
the Danish 1.5m telescope (D1.5m), the 1m telescope at the Siding Spring Observatory
(SSO), VLT/FORS1, and VLT/FORS2. The $R$-band data by the Watcher telescope were
processed and made public for the first time, which not only are consistent with other
$R$-band data but provide the accurate peak time, 0.3 days since the BAT trigger, of the
$R$-band afterglow lightcurve. We have applied the correction for the Galactic
extinction, $E(B-V)=0.057$ mag (\citealt{Schlegel98}), and subtracted the contribution of
host galaxy, $R_{\rm host}=22.46\pm0.04$ (\citealt{DellaValle06}).

We collected {\it Swift}/UVOT data from \cite{Mangano07}. For the UVOT bands,
we also applied the correction for Galactic extinction and subtracted the
contribution of host galaxy using the template in \cite{Mangano07}.

The {\it Swift}/XRT lightcurve and spectrum data were collected from the UK
Swift Science Data Centre (\citealt{Evans07}). The X-ray lightcurves at 0.3 keV
and 1.5 keV are shown in Fig.~\ref{0614LC} so that the spectral slope,
$\beta_{X}\sim0.89$, in the 0.3-10 keV band is taken into account when we
performed the numerical fitting.

The {\it Swift}/BAT lightcurve in the 15-350 keV band was processed with the
{\tt batgrbproduct} task of the HEAsoft 6.5.1.

\section{Interpretation of the two afterglows}

Suppose the radial density profile of the circumburst medium takes the form $n (r)
\propto r^{\,-k}$, then $k=0$ if the medium is interstellar medium-like (ISM-like) while
$k=2$ if the medium stellar wind-like (WIND-like).

We use the standard fireball afterglow theory reviewed by, e.g., \cite{Piran05}, with the
simple microphysical assumptions of constant energy fractions imparted to the swept-up
electrons, $\epsilon_e$, and to the generated magnetic field, $\epsilon_B$, respectively.
For the evolution of the synchrotron spectrum, we adopt the prefactors of equation (1) in
\cite{Yost03} for the ISM scenario, and those of equations (11-14) in \cite{CL00} for the
WIND scenario. We note that both cases lead to the afterglow closure relations made of
the temporal decay index $\alpha$ and the spectral index $\beta$, depending upon the
spectral segment and the electron energy distribution index, $p$ (see Tables II and IV of
\citealt{Piran05}).

For numerical calculation, we follow the general treatment of \cite{Huang00} and
\cite{FP06}, that is, one first calculates the overall dynamical evolution of the GRB
outflow, and then calculates the synchrotron radiation at different times, including
different corrections such as, e.g., the equal-arrival-time-surface effect and the
synchrotron-self-absorption effect.

\subsection{Constraints on GRB\,060505} 
\begin{figure}
\centerline{\includegraphics[width=9cm, height=6cm,angle=0]{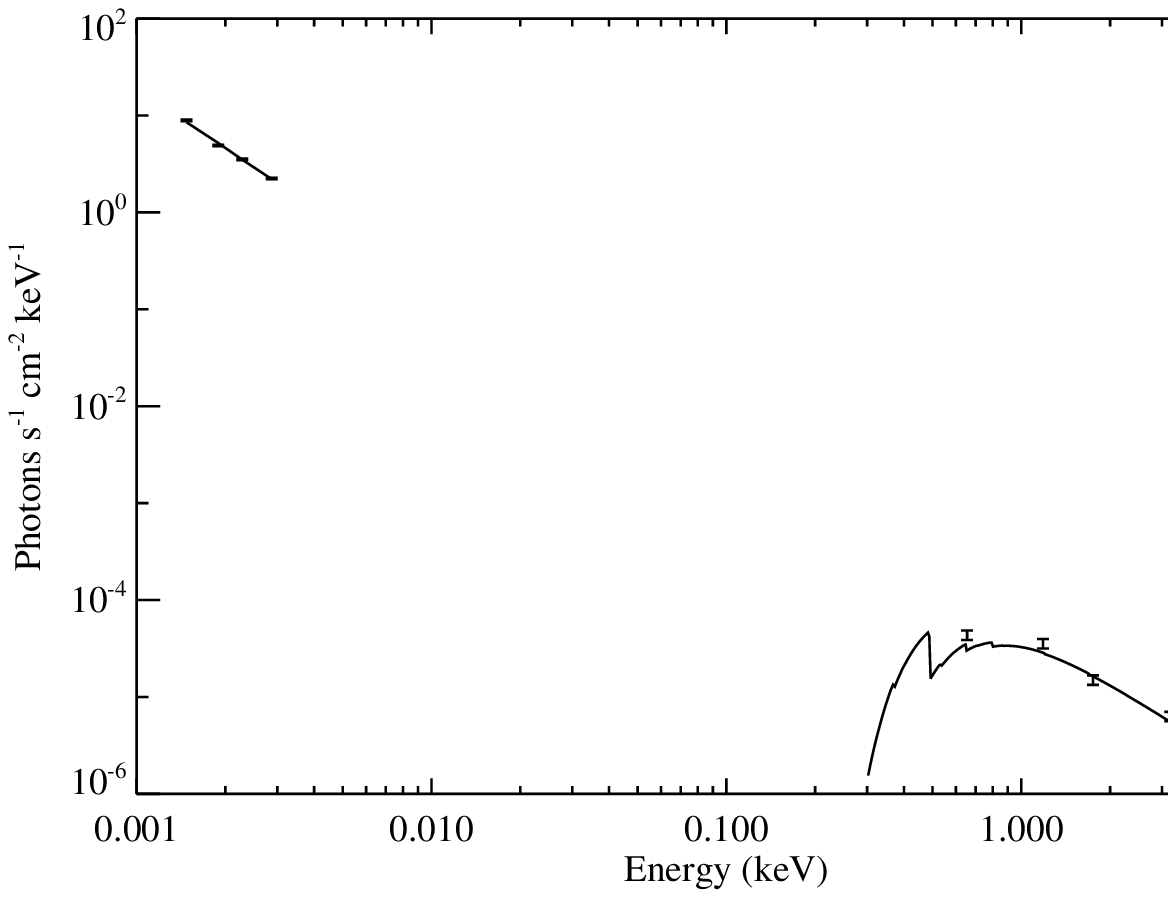}} \caption{
The broadband SED for GRB\,060505 at the epoch of the multi-band optical observation
(1.125 days after the burst). The SED can be fitted with a single absorbed power-law with
slope $\beta = 0.97\pm0.03$. \label{GRB060505SED}}
\end{figure}

To establish a broadband spectral energy
distribution (SED) we extrapolate the X-ray flux to the epoch of the optical data (i.e.,
1.125 days after the burst). The optical data were corrected for foreground Galactic
extinction of $A_B = 0.089$, $A_V=0.068$, $A_R=0.055$, $A_I=0.040$ \citep{Schlegel98}. We
fit the SED in count space (see \citealt{Starling}) with an absorbed power law model,
where Galactic absorption in the X-rays is fixed at $N_{\rm H} = 1.8\times 10^{20}$
cm$^{-2}$ \citep{Kalberla} and intrinsic X-ray absorption and optical/UV extinction in
the host galaxy are free parameters. Solar metallicity is assumed in the X-ray absorption
model, and extinction in the host is assumed to be SMC-like \citep{Pei}. The resulting
SED is shown in Fig.~\ref{GRB060505SED}. The derived spectral slope is $\beta_{OX} =
0.97\pm0.03$. The host galaxy extinction is consistent with zero, but with a best fitting
value of $E(B-V)=0.015$ mag and a 90\% upper limit of $E(B-V)=0.05$ mag, while the X-ray
absorption is found to be $N_{\rm H} = (0.2^{+0.2}_{-0.1}) \times 10^{22}$~cm$^{-2}$
(where errors are quoted at the 90 \% confidence level). The fit has a $\chi^2$/dof of
5.3/4.

\begin{figure}
\centerline{\includegraphics[width=10cm,height=10cm,angle=0]{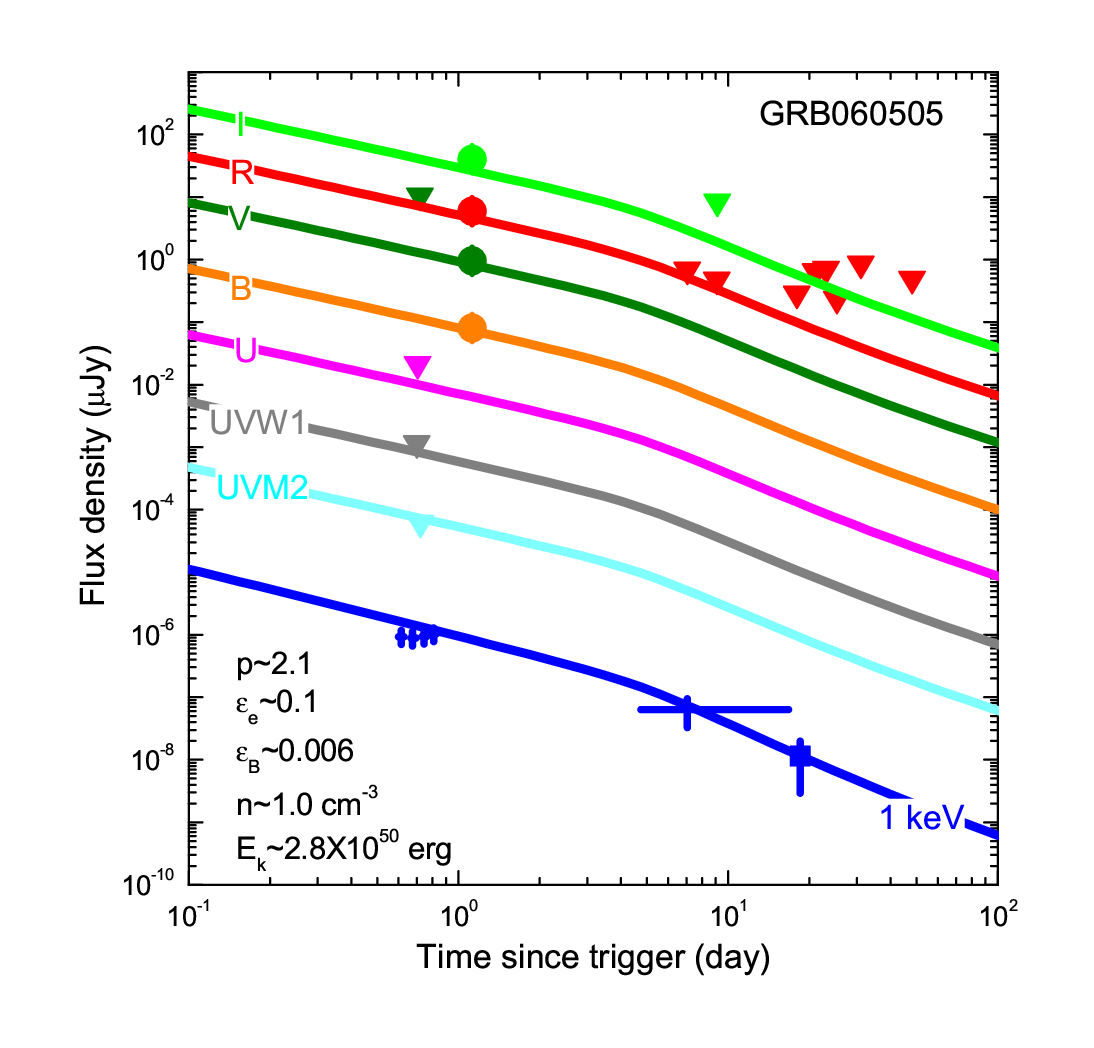}}
\caption{The multi-band lightcurves for the afterglow of GRB\,060505. For clarity, the
shown flux densities in the I, R, V, B, U, UVW1, UVM2, and 1 keV bandpass are 5, 1, 1/50,
1/500, 1/5000, 1/50000, 1/50000 times that of their real flux densities, respectively.
Points and crosses represent the measurements with errorbars while triangles represent
upper limits. Also marked are the parameters used for a good fit.\label{0505LC}}
\end{figure}

The index $\beta_{OX} = 0.97\pm0.03$ indicates that the cooling frequency
$\nu_c$ is already below the optical at this time and the energy spectral index
$p$ is slightly larger than 2. Indeed, the numerical fit, shown in
Fig.~\ref{0505LC}, yields $p\sim2.1$, and other afterglow parameters are
$n\sim1\,{\rm cm^{-3}}$, $\epsilon_e\sim0.1$, $\epsilon_B\sim0.006$,
$E_k\sim2.8\times10^{50}\,{\rm erg}$, and the half-opening jet angle $\theta_j
\sim0.4\,{\rm rad}$. These parameters are within the range of LGRB afterglows,
but the limited data prevent us from getting more insight on the properties of
this afterglow. Both the SED measurement and the numerical fit tend to render it
unnecessary to employ {\it ad hoc} models (e.g., macronova in \citealt{Ofek07})
to interpret this afterglow.

\subsection{Preliminary constraints on GRB\,060614} 
\begin{figure}
\centerline{\includegraphics[width=6cm, height=9cm,angle=90]{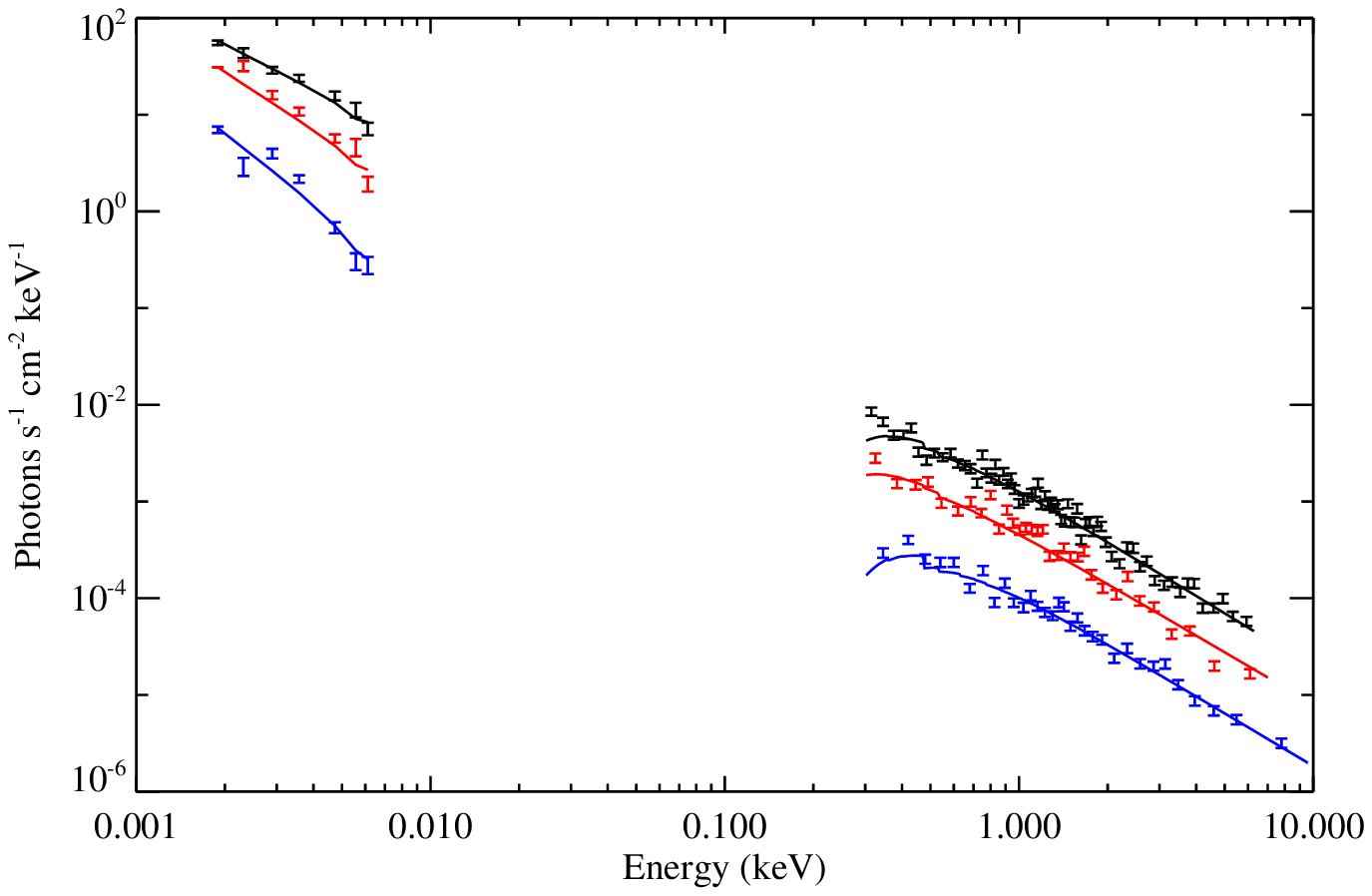}}
\caption{The broadband SEDs for GRB\,060614 at the epochs of 0.187, 0.798, and
1.905 days from top to bottom. A broken power-law is only required during the first epoch.
Refer to Table~\ref{0614SED} for detailed measurements. \label{GRB060614SED}}
\end{figure}

\begin{figure*}
\centerline{\includegraphics[width=18cm,height=16cm]{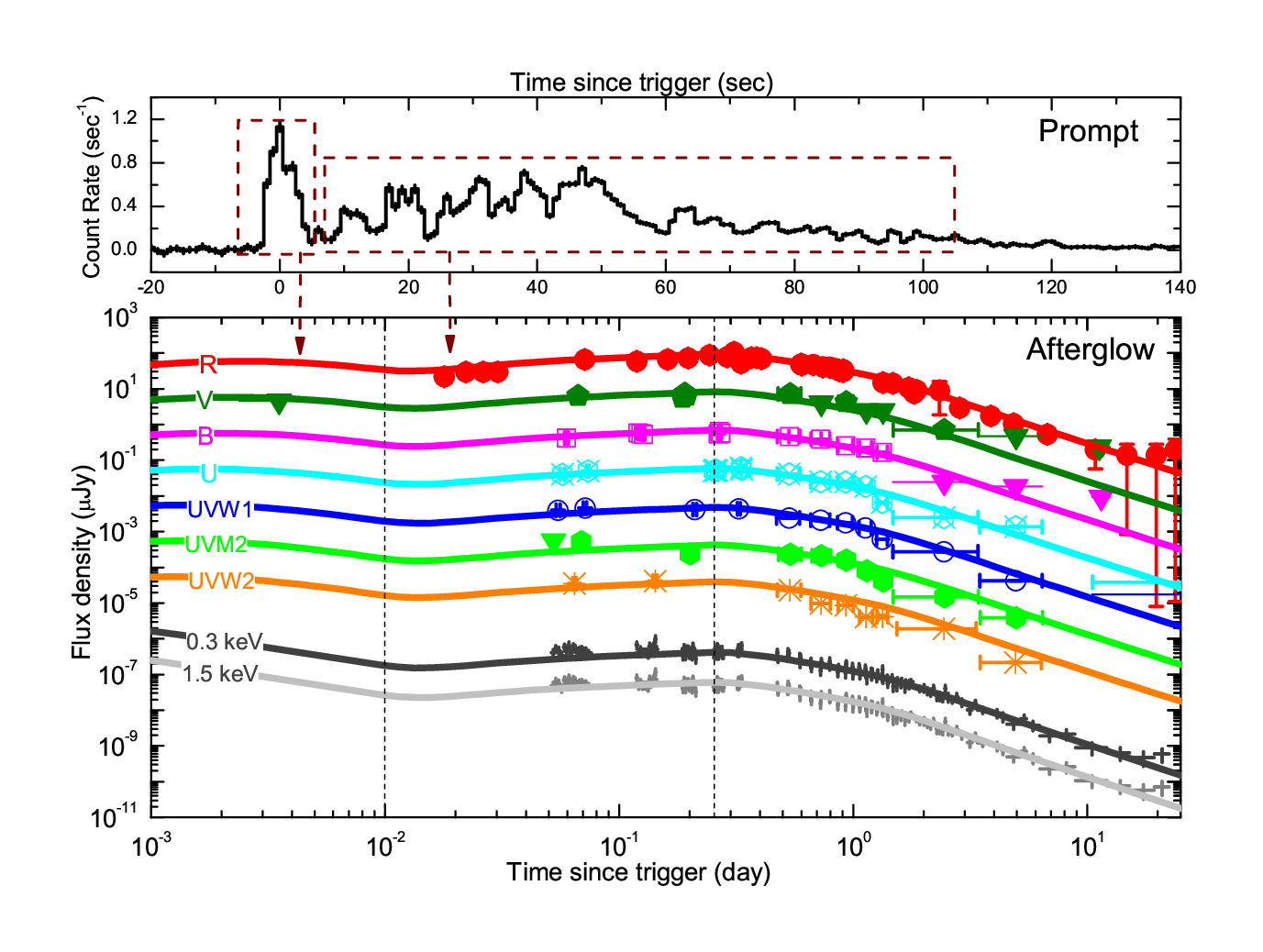}} \caption{The
temporal lightcurves for the prompt phase (upper panel) and for the afterglow (lower
panel) of GRB\,060614. {\it Upper:} the prompt lightcurve in the 15-350 keV band,
consisting of a hard spike of duration $\sim5$~s and a soft tail of $\sim100$~s. {\it
Lower:} numerical fit to the afterglow lightcurves which consists of an episode of energy
injection enclosed by two vertical dashed lines. For clarity, the shown flux densities in
the R, V, B, U, UVW1, UVM2, UVW2, 0.3 keV, and 1.5 keV bands are $10^0$, $10^{-1}$,
$10^{-2}$, $10^{-3}$, $10^{-4}$, $10^{-5}$, $10^{-6}$, $2\times10^{-7}$, $10^{-7}$ times
that of their real flux densities, respectively. In our model, the energy injection
corresponds to the soft tail in the prompt phase while the main afterglow corresponds to
the hard spike$--$this correlation is illustrated by two arrows from the upper prompt
panel to the lower afterglow panel. \label{0614LC}}
\end{figure*}

We constructed afterglow SEDs in
GRB\,060614 at three epochs, i.e., 0.187 days, 0.798 days, and 1.905 days. The results
are listed in Table~\ref{sedresults} and shown in Fig.~\ref{GRB060614SED}. Our SEDs are
fully consistent with those in \cite{Mangano07} measured at 0.116, 0.347, 0.694, and
1.736 days. Both works show that there exists a spectral break between the optical and
the X-ray before $\sim$0.26 days while afterwards both the optical and the X-ray are in
the same spectral segment with the spectral index $\beta_{OX}\sim0.8$.

The afterglow lightcurves show that energy injection exists between $\sim 0.01$ and
$\sim0.26$ days, which presumably would change the SED during this period. If the
injection frequency is between the optical and the X-ray, then it may lead to that the
lightcurves have a very shallow rising in the optical band and a very shallow decay in
the X-ray band. Indeed, the observational data fit this interpretation very well
(\citealt{Mangano07}). We found that this optical-to-Xray lightcurve behavior still holds
after performing different corrections. The result of our analysis shows that a flat or
gradually increasing lightcurve is generally a description as good as a slow decaying
except in the X-ray band. A temporal peak, clearly shown in optical bands, exists $\sim
0.3$ days after the burst. Afterwards the afterglow decays with $\alpha_1\sim1.1$ until
$t_b\sim1.4$ days when it steepens significantly to $\alpha_2\sim2.5$. There is only one
$V$-band upper limit before $\sim 0.01$ days, the starting time of energy injection.
Therefore, to constrain the afterglow properties we use data after $\sim0.26$ days.

The index $\beta_{OX}\sim 0.8$ after $\sim0.26$ days indicates that the afterglow is
in the slow cooling case of \[ \nu _a < \nu _m  < \nu _{O}  < \nu _X  < \nu _c
\] until at least a few days. Using the afterglow closure relations we find
$\beta=(p-1)/2\sim0.8$, which gives the energy spectral index $p\sim2.6$, which then
gives the decay index $\alpha=3(p-1)/4\sim1.2$, in good agreement with the
observed decay law ($\alpha_1\sim1.1$). Furthermore, this $p$ value of $\sim
2.6$ is very close to the observed decay index of $2.5$ after $t_b \sim 1.4$
days, in good agreement with the theoretically predicted decay law after a jet
break, i.e., $\alpha \sim p$ at any wavelength from X-ray to optical. Therefore,
as discussed by \citealt{Mangano07},
the break at $t\sim1.4$ days is likely the so-called {\it jet break}
(\citealt{Rhoads99,Sari99,Zhang06,Nousek06}).

Since the optical decay is never faster than the X-ray decay, there is no indication of a
WIND-like circumburst medium for this afterglow. Also since radio data are not available,
analytical constraint on $\nu_a$ is impossible for this afterglow. Using the earliest
useful data, we put a lower limit on $F_{\nu}^{max}$ and an upper limit on $\nu_m$ as
\begin{equation}
F_{\nu}^{max}= 1.6 \, (z+1) \,D_{28}^{-2}\, \epsilon_{B,-2}^{0.5}\, E_{52} \,n^{0.5}
> 105.5\times 10^{-3}\,\,\,\,{\rm mJy}
\end{equation}
and
\begin{equation}
\nu_m =  3.3\times10^{14} \,  (z+1)^{0.5}\, \epsilon_{B,-2}^{0.5}
\,\bar{\epsilon}_e^{2}\, E_{52}^{0.5}\, t_d^{-1.5} <4.69 \times 10^{14}\,\,\,\,{\rm Hz}
\end{equation}
where $t_d=0.3$ and $z=0.125$. Using the latest useful data, we put the lower limit on
$\nu_c$ as
\begin{equation}
\nu_c  =  6.3\times10^{15}  \, (z+1)^{-0.5}\, \epsilon_{B,-2}^{-1.5} \,E_{52}^{-0.5}
\,n^{-1} \,t_d^{-0.5}> 10^{18}\,\,\,\,{\rm Hz}
\end{equation}
where $t_d=10$. In detail, the above three equations give rise to the constraints
\begin{equation}
\begin{array}{lcc}
\epsilon_{B, - 2}^{0.5}\, {\rm{ }}E_{52}\, {\rm{ }}n > 0.0019\\
\varepsilon _{B, - 2}^{0.5}\, {\rm{ }}\varepsilon _e^2\, {\rm{ }}E_{52}^{0.5}  < 1.564\\
\epsilon_{B, - 2}^{ - 1.5}\, {\rm{ }}E_{52}^{ - 0.5}\, {\rm{ }}n^{ - 1}  > 532.4.\\
\end{array}
\end{equation}

\subsection{Numerical constraints on GRB\,060614} We find that the afterglow data can be
reasonably reproduced by the following parameters (see Fig. \ref{0614LC} for a plot of
our fit): $p\sim 2.5$, $\epsilon_e \sim 0.12$, $\epsilon_B\sim 0.0002$, $E_{\rm k}\sim
6\times 10^{50}$ erg, $n\sim 0.04~{\rm cm^{-3}}$, and $\theta_j \sim 0.08$ rad. The
energy injection takes place at $t_i\sim 8\times10^2$~s and ends at $t_e\sim 2\times
10^4$~s after the burst, and the energy injection is nearly a constant with a rate
$L_{\rm inj}\sim 1.2\times10^{48}~{\rm erg~s^{-1}}$. Substituting these parameters into
equation (8), the numerical constraint is in agreement with the analytical constraint.

If the energy injection is from the wind of a millisecond magnetar, to fit the
observational data at the late stage of the whole energy injection period, the magnetar
is required to have dipole radiation $L_{\rm dip}(t)\approx {2.6\times
10^{48}/(1+z)}~{\rm erg~s^{-1}}~B_{\bot,14}^2R_{s,6}^6\Omega_4^4[1+{t/((1+z)T_o)}]^{-2}$,
where $B_{\bot}$ is the dipole magnetic field strength of the magnetar, $R_s$ is the
radius of the magnetar, $\Omega$ is the initial angular frequency of radiation,
$T_o=1.6\times 10^4 B_{\bot,14}^{-2} \Omega_4^{-2}I_{45}R_{s,6}^{-6}$~s is the initial
spin-down timescale of the magnetar, and $I\sim 10^{45}~{\rm g~cm^2}$ is the typical
moment of inertia of the magnetar (\citealt{Pacini67}). However, because the optical flux
is roughly proportional to $\bar{E}_{\rm k}$, where $\bar{E}_{\rm k}$ is the sum of the
isotropic-equivalent kinetic energy $E_{\rm k}$ and the injected energy, then the
predicted optical flux at the early stage of the energy injection period (e.g.,
$t\sim0.02$ days in Fig. 5) would be much higher than the observed flux. In detail, at
$t\sim 0.02$ day, there would be $\bar{E}_{\rm k}\sim E_{\rm k}+L_{\rm inj}t \sim 3
E_{\rm k}$, indicating that the predicted optical flux should be $\sim$3 times the
observed flux. Therefore the magnetar model is not convincing.

Note that the prompt $\gamma-$ray lightcurve may have two components: the earlier hard
spike with an isotropic energy $E_{\gamma,h}\sim 3.7\times 10^{50}$ erg and the latter
soft tail (sometimes called extended emission) with an isotropic energy $E_{\gamma,s}
\sim 1.7\times 10^{51}$ erg. The early part is spectrally hard thus the outflow might be
ultra-relativistic, while the latter part is spectrally soft suggesting the bulk Lorentz
factor of the outflow became lower. This is because the optically thin condition yields a
lower limit on $\Gamma \geq 20 (L_{\rm outflow}/10^{50}~{\rm erg~s^{-1}}) ^{1/5} \delta
t^{-1/5}$, where $L_{\rm outflow}$ is the total luminosity of the outflow, and $\delta t$
is the typical variability timescale of the late soft $\gamma-$ray emission
(\citealt{RM94}). In our numerical calculation, we find the bulk Lorentz factor of the
forward shock $\Gamma \sim 26$ at $t\sim 10^3$~s while $\Gamma \sim 16$ at $t\sim 2\times
10^4$~s. If the energy carried by the material of the late time GRB ejecta satisfies the
relation $E(>\Gamma)\propto \Gamma^{-5}$ (\citealt{RM98}) for $16<\Gamma<26$, the
constant energy injection form taken in the afterglow modeling can be reproduced
(\citealt{Zhang06}). In this model, for the outflow accounting for the hard spike
emission the energy efficiency is $\sim E_{\gamma,h}/(E_{\gamma,h}+E_{\rm k})\sim40\%$,
while for the outflow accounting for the soft tail emission, the energy efficiency is
$\sim E_{\gamma,s}/[L_{\rm inj}(t_e-t_i)+E_{\gamma,s}]\sim 8\%$. The decreasing
efficiencies from early spike to late tail may be due to the smaller contrast between the
Lorentz factors of the fast material and that of the slow material.

\section{Discussion and conclusion}
As shown in several early papers, for GRB\,060505 (\citealt{Fynb06}) and GRB\,060614
(\citealt{Fynb06,DellaValle06,Gal-Yam06}) there is no accompanying SN emission, down to
limits hundreds of times fainter than the archetypical SN\,1998bw that accompanied
GRB\,980425, and fainter than any Type Ic SN ever observed. Multi-wavelength observations
of the early afterglow exclude the possibility of significant dust obscuration. For
GRB\,060505 the properties of the host galaxy (\citealt{Ofek07,Thoene08}) as well as the
spectral lag of the prompt emission (\citealt{McBreen08}) is most consistent with the
properties expected for the long-duration class of GRBs. For GRB\,060614 the duration of
the prompt emission places the burst firmly within the long-duration class of GRBs, but
the negligible spectral lag (\citealt{Gehrels06}) and the relatively modest
star-formation activity of the host galaxy is more similar to the expected properties for
the short-duration class of GRBs. In this paper we have investigated whether or not the
properties of the two afterglows could provide some hints to the most likely progenitor
types for these bursts.

For GRB\,060505, the numerical fit of its afterglow shows that the standard jetted
external shock wave model is consistent with the data, yielding a typical interstellar
medium density of $n \sim 1\,{\rm{cm}}^{-3}$, a wide jet angle $\theta_j \sim25^\circ$,
and a possible jet break at $t_b \sim3$ days. For GRB\,060614, the standard external
shock wave model is again consistent with the data, but apparently needing to invoke
energy injection. The afterglow shows the clearest achromatic jet break among all
{\swift} bursts studied so far, decaying in a broken power-law from $\alpha_1\sim-1.1$ to
$\alpha_2\sim -2.5$ at $t_b\sim 1.4$ days. An achromatic peak, especially in the {\it
UBVR} bands, occurs at $\sim 0.3$ days, which we interpret as resulting from an episode
of strong energy injection as \cite{Mangano07} suggested. Numerical fit yields a
circumburst density of $n \sim 0.04 \,{\rm{cm}}^{-3}$, being consistent with the inferred
value in \cite{caito08},  and a jet angle of $\theta_j\sim 5^\circ$. The inferred
afterglow parameters for the two bursts fall within the range for other LGRBs.  If it had
not been for the observed absence of associated SNe we would have no reason, from the
afterglow properties, to question their classification as LGRBs.

After discovery of these two GRBs, to reconcile all SN-observed and SN-less GRBs within
the conventional framework of short ($\lesssim 2$ s) and long ($\gtrsim 2$ s) GRBs, a new
classification was proposed, in which GRBs featuring a short-hard spike and a (possible)
long-soft tail would be ascribed to the conventional short class, or Type I in the new
taxonomy, while all other GRBs, or Type II, would comprise the conventional long class
(\citealt{Zhang07}, see also \citealt{Kann07}). The new classification expands the range
of the conventional short class, and is applicable to GRB\,060505 and GRB\,060614, which
then would be SN-less due to a merger-related progenitor rather than SN-less massive
stellar death.

\begin{figure*}
\centerline{\includegraphics[width=19cm,height=12cm]{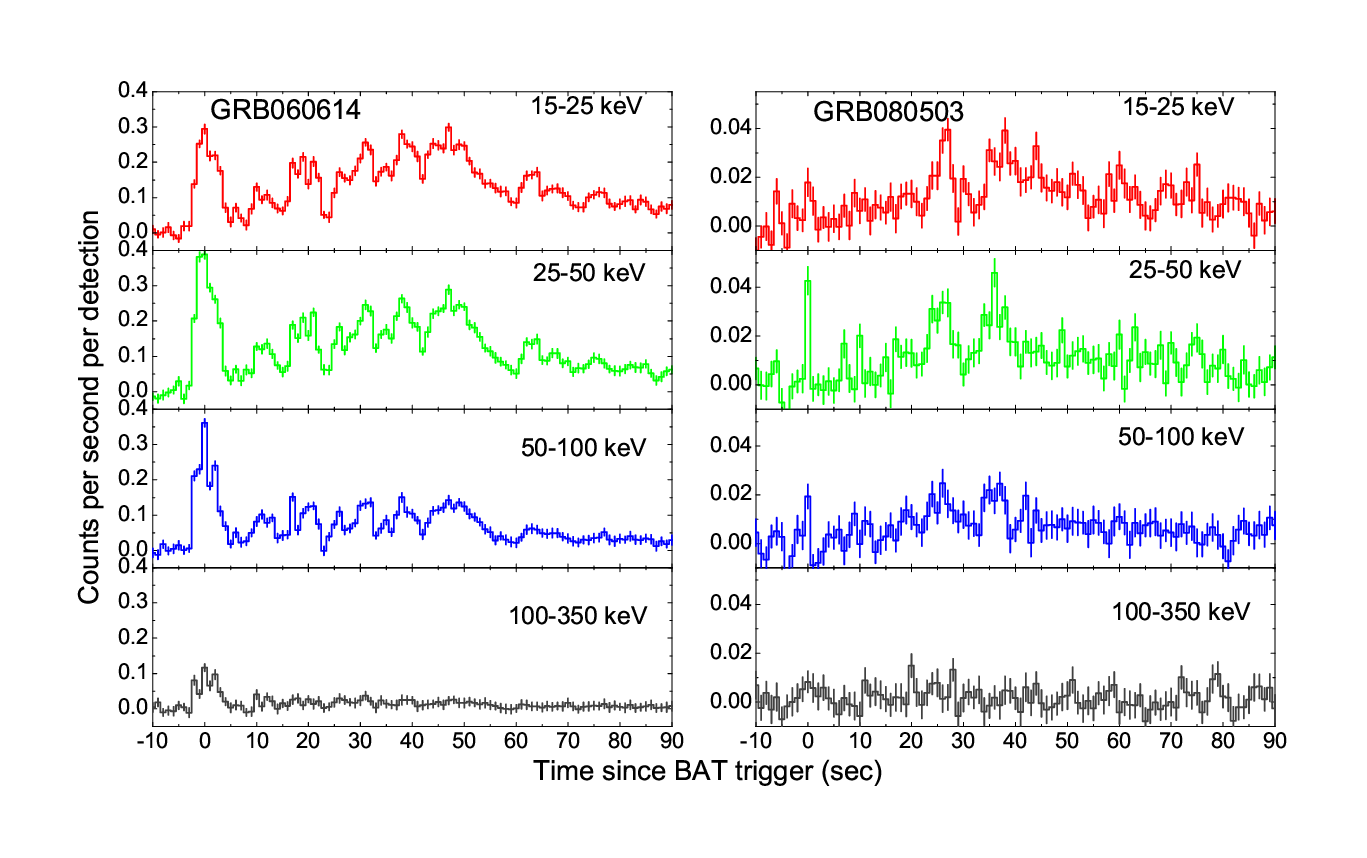}} \caption{Comparison
of the prompt lightcurves (1~s binning) of GRB\,060614 (left column) and GRB\,080503
(right column) detected by BAT in different energy bins. Morphologically both bursts
consist of a spike emission followed by an extended emission, namely, a tail emission.
For each energy bin of either burst, comparing the count rates in the spike phase and in
the tail phase yields a rough estimate of the spectral hardness. In GRB\,060614 the tail
is considerably softer than the spike, while in GRB\,080503 the tail is comparably as
hard as the spike, being consistent with the comparison of spectral measurements in
Table~\ref{080503}. The spectral lags for both spikes are consistent with zero; the lag
for the tail of GRB\,060614 is consistent with zero while the lags for the tail of
GRB\,080503 are $0.8^{+0.3}_{-0.4}$ s for the 25-50 keV vs. 15-25 keV band and
$0.8^{+0.4}_{-0.5}$ s for the 50-100 keV vs. 15-25 keV band. \label{comparison}}
\end{figure*}

However, a recent burst, GRB\,080503, seems to challenge the new classification. This
burst also has a temporal spike $+$ tail structure in the prompt emission phase. The
$T_{\rm 90}$ values of the initial spike and the total emission in the 15-150 keV band
are $0.32\pm0.07$~s and $232$~s, respectively (\citealt{Perley08}). The fluence of the
non-spike emission measured from 5~s to 140~s after the BAT trigger in the 15-150 keV
band is $(1.86\pm 0.14)\times 10^{-6}\,{\rm erg\,cm^{-2}}$, being around 30 times that of
the spike emission in the same band. This fluence ratio is much higher than the ratio of
around 6 for GRB\,060614, and higher than any previous similar {\it Swift} GRB. For
GRB\,060614 and GRB\,080503 we extracted the BAT lightcurves in different energy bins,
shown in Fig.~\ref{comparison}, for comparison study. For GRB\,080503 we have analysed
the spectral evolution during the prompt emission period by BAT and XRT and list our
analysis and that of other groups in Table \ref{080503}. From these we conclude that: (1)
The results of different groups are fully consistent with each other. (2) The photon
indices for the spike and non-spike emissions are consistent within their error regions
($90\%$ confidence level). A strong spectral softening from the spike phase to the
non-spike phase can be excluded. A cutoff power-law fit does not improve the fitting,
yielding the error of $E_{\rm peak}$ larger than 100\%. (3) During the BAT-XRT overlap
period, the XRT spectra are always harder than the BAT spectra, which implies that the
BAT$+$XRT spectra (0.3-150 keV) would be harder than the BAT spectra (15-150 keV) alone.
This adds more evidence that from spike to non-spike the spectra did not soften
considerably. In addition, note that for all conventional LGRBs (e.g., from BATSE), there
is a general trend that the spectra during the prompt emission would soften mildly from
the beginning time to the ending time (\citealt{Norris00}). Therefore, the prompt
spectral evolution of GRB\,080503 is different from that of GRB\,060614.

\begin{figure*}
\centerline{\includegraphics[width=16cm,height=13cm]{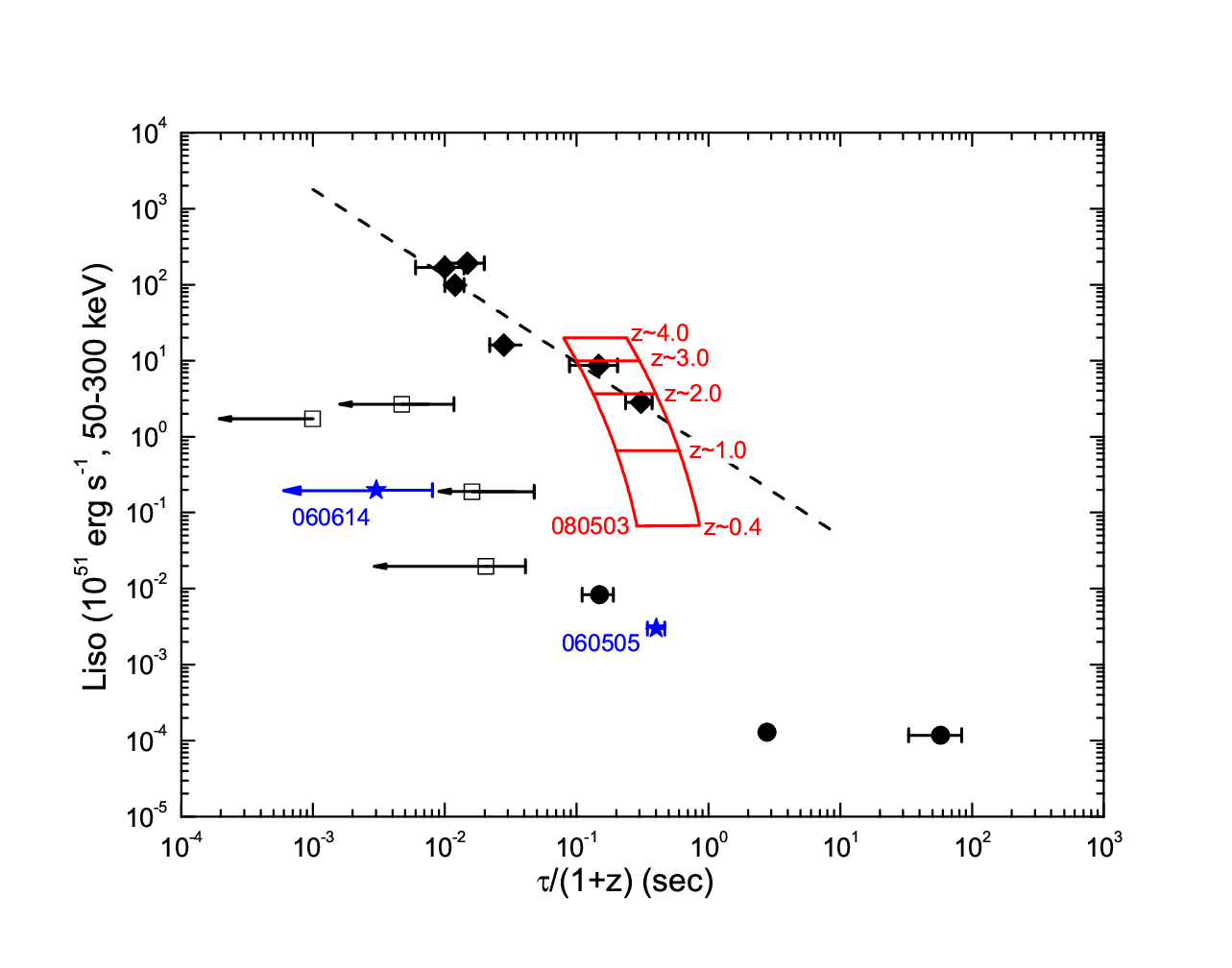}} \caption{Spectral lag-peak
luminosity relation for GRBs. The Norris LGRB data (diamonds) and the associated fit
(dashed line) are from \cite{Norris00}. The data of SGRBs (open squares), the nearby
SN-bright GRBs 980425, 031203, and 060218 (filled circles), the nearby SN-less GRBs
060505 and 060614 (stars) are from \cite{McBreen08} and references therein. Also shown is
the spectral lag-peak luminosity region for the extended emission of GRB 080503 (closed
region) due to lacking the spectroscopic redshift of this burst. The position of this
burst moves upward in the plane as its {\it assumed} redshift increases. \label{SL}}
\end{figure*}

In addition, we computed the spectral lags in different energy bands using the 64 ms
binning lightcurves, following the method in \cite{Norris00} and \cite{Chen05}. The lag
during the initial spike phase is consistent with zero. From 5 to 50 s since BAT trigger,
the lags are $0.8^{+0.3}_{-0.4}$ s for the 25-50 keV vs. 15-25 keV band and
$0.8^{+0.4}_{-0.5}$ s for the 50-100 keV vs. 15-25 keV band respectively, both well above
the lag range for SGRBs. Again the lag of GRB\,080503 is in contrast with that of
GRB\,060614. The redshift of GRB\,080503 was not measured mainly because its optical and
X-ray afterglows became very faint shortly after the BAT trigger (never exceeding 25 mag
in deep observations starting at $\sim 1$ hr since trigger), but the g-band photometric
detection imposes a limit of approximately z$<$4 (\citealt{Perley08}). In Fig.~\ref{SL}
we show the possible position of GRB\,080503 in the spectral lag-peak luminosity plane
relative to the positions of previous LGRBs and SGRBs. As can be seen, for the extended
emission of GRB\,080503, its position is outside the SGRB population at a very high
confidence level regardless of its redshfit. For the spike emission of GRB\,080503, its
position is within the SGRB population because of a peak luminosity comparable to that of
the extended emission, as well as a negligible spectral lag.

The very faint optical and X-ray afterglows of GRB\,080503 may indicate the circumburst
density is very low. This is consistent with the fact that the afterglow is located away
from any host galaxy down to 28.5 mag in deep \textit{Hubble Space Telescope} imaging
(\citealt{Perley08}). These observational signatures contribute to put GRB\,080503 into
the merger class. If we ignore the lag function in classifying GRBs and relax the
restriction of ``soft-tail'' for a Type I GRB to ``either soft- or hard- tail'', then it
would be (even more) ambiguous, also operationally difficult to define the type of
spike$+$tail GRBs among all long GRBs, especially among those at $z\gtrsim0.7$ for which
a SN search in their afterglows is difficult or impossible. Bearing in mind that GRBs can
be either luminous or under-luminous, and the redshift could be either high or low, then
the problems would be: up to what duration should be classified as a spike, and how much
should the flux ratio be for the spike component over the non-spike component? If
GRB\,080503 is interpreted as a merger burst, then it enhances the possibility that a
merger could produce a long GRB, at least in the prompt emission phase, mimicking the one
produced by a collapsar. GRB\,080503 is a dark burst with the optical-to-Xray spectral
slope $\beta_{OX}$ well below 0.5, the critical value for defining a dark burst
(\citealt{Jakobsson04}), at 0.05 days after the burst. We speculate that some other dark
bursts may have their progenitor same as GRB\,080503.

The progenitors of GRB\,060505 and GRB\,060614 remain uncertain based on their
observations and the current GRB and SN theory. Other than the lack of a SN component,
their afterglows are actually not peculiar when compared with the afterglows of other
LGRBs. According to the current theoretical study, in the core-collapse scenario the
``fallback''-formed black holes or progenitors with relatively low angular momentum could
produce such SN-less GRBs (e.g.,
\citealt{Frye06,Sumiyoshi06,Nakazato08,Kochanek2008,Valenti09} for observational
existence of an extremely faint Ibc SN), or in the merger scenario the two compact
objects also could produce such SN-less GRBs if the formed remnant, a differentially
rotating neutron star or an uniformly rotating magnetar, has not collapsed into a black
hole immediately \citep{KR98,RRD03}. A difference existing in the afterglows between
these two scenarios is that the collapsar model predicts a WIND-like circumburst medium
created by the Wolf-Rayet progenitor star while the merger model does not. As a matter of
fact the WIND signature is not clearly evident in most LGRBs, but this should not
necessarily lead to the merger origin for these bursts because the definite WIND
signature is an ideal case for a constant wind off a massive star. If lucky enough, the
core-collapse origin for a SN-less GRB (no matter whether it has the hard-spike $+$
soft-tail structure) will be quite certain if in the afterglow, either the X-ray flux
$F(t)$ decays as $\propto t^{-\alpha}$ with its spectrum as $\propto
\nu^{-(2\alpha-1)/3}$ for the $\nu_m < \nu_{X} < \nu_c$ phase (normally in early
afterglow), or the optical decay index is larger than the X-ray decay index by a factor
of $\sim1/4$ for the slow cooling phase of the WIND scenario.

\acknowledgements We thank the Paranal staff for performing the VLT observations reported
in this paper. The Dark Cosmology Centre is funded by the Danish National Research
Foundation. DX acknowledges funding from the European Commission under the Marie Curie
Host Fellowships Action for Early Stage Research Training SPARTAN programme (Centre of
Excellence for Space, Planetary and Astrophysics Research Training and Networking)
Contract No. MEST-CT-2004-007512, University of Leicester, UK. DX thanks Yizhong Fan for
stimulating discussion, Kim Page for help in {\it Swift} data reduction, Li Chen for help
in the spectral lag computation, and Daniele Malesani for comments on the manuscript. We
thank the referee for valuable comments and suggestions. JS is a Royal Swedish Academy of
Sciences Research Fellow supported by a grant from the Knut and Alice Wallenberg
Foundation. This work made use of data supplied by the UK Swift Science Data Centre at
the University of Leicester.

\clearpage
\clearpage
\clearpage
\clearpage
\clearpage
\clearpage
\clearpage
\clearpage

\begin{deluxetable}{cccccc}
\tablecolumns{6} \tablewidth{0pt}

\tablecaption{Log of optical observations of GRB\,060505.}

\tablehead{
\colhead{Time$^a$} & \colhead{Bandpass} & \colhead{Vega Mag$^b$} & \colhead{$\sigma$(mag)} & \colhead{Instrument} & \colhead{Ref}\\
\colhead{[day]} & \colhead{} & \colhead{} & \colhead{} & \colhead{} & \colhead{}\\}

\startdata
0.702    &\ UVW1 &\ $>$20.71  &\ --   &\ UVOT        &\ 1\\
0.706    &\  U   &\ $>$20.31  &\ --   &\ UVOT        &\ 1\\
0.721    &\  V   &\ $>$20.34  &\ --   &\ UVOT        &\ 1\\
0.724    &\ UVM2 &\ $>$22.05  &\ --   &\ UVOT        &\ 1\\
1.102    &\  r   &\  21.65    &\ 0.16 &\ GMOS        &\ 1\\
1.118    &\  g   &\  22.37    &\ 0.08 &\ GMOS        &\ 1\\
1.125    &\  I   &\  21.21    &\ 0.04 &\ VLT+FORS1   &\ 2\\
1.125    &\  R   &\  21.74    &\ 0.04 &\ VLT+FORS1   &\ 2\\
1.125    &\  V   &\  22.14    &\ 0.04 &\ VLT+FORS1   &\ 2\\
1.125    &\  B   &\  22.48    &\ 0.04 &\ VLT+FORS1   &\ 2\\
7.041    &\  g   &\  $>$24.74 &\ --   &\ GMOS        &\ 1\\
7.055    &\  r   &\  $>$24.02 &\ --   &\ GMOS        &\ 1\\
9.078    &\  g   &\  $>$24.54 &\ --   &\ GMOS        &\ 1\\
9.092    &\  r   &\  $>$24.32 &\ --   &\ GMOS        &\ 1\\
9.105    &\  i   &\  $>$22.96 &\ --   &\ GMOS        &\ 1\\
18.025   &\  R   &\  $>$24.95 &\ --   &\ VLT+FORS2   &\ 3\\
21.125   &\  R   &\  $>$24.05 &\ --   &\ D1.5m+DFOSC &\ 3\\
23.125   &\  R   &\  $>$23.95 &\ --   &\ D1.5m+DFOSC &\ 3\\
25.325   &\  R   &\  $>$25.15 &\ --   &\ Keck+LRIS   &\ 3\\
31.125   &\  R   &\  $>$23.75 &\ --   &\ D1.5m+DFOSC &\ 3\\
48.125   &\  R   &\  $>$24.35 &\ --   &\ D1.5m+DFOSC &\ 3\\
\hline
\enddata
\tablenotetext{a}{Time since BAT trigger.}

\tablenotetext{b}{The Vega magnitude is after correction for the Galactic extinction of
$E(B-V)=0.02$ mag, and image subtraction to remove the host contribution.}

\tablerefs{(1) \citealt{Ofek07}; (2) This work; (3) \citealt{Fynb06}.}

{\label{0505Optical}}
\end{deluxetable}
\clearpage

\begin{deluxetable}{ccccc}
\tablecolumns{4} \tablewidth{0pt} \tablecaption{Log of $R$-band optical observations of
GRB\,060614} \tablehead{
\colhead{Date$^a$} & \colhead{Veg Mag$^b$} & \colhead{$\sigma$(mag)} &\ Instrument & \colhead{Ref}\\
\colhead{[days]} & \colhead{} & \colhead{}& \colhead{} & \colhead{}}

\startdata

0.67347 &\ 19.45 &\ 0.01 &\ D1.5m+DFOSC &\ 1\\
0.74059 &\ 19.60 &\ 0.01 &\ D1.5m+DFOSC &\ 1\\
0.79292 &\ 19.64 &\ 0.01 &\ D1.5m+DFOSC &\ 1\\
0.84034 &\ 19.73 &\ 0.01 &\ D1.5m+DFOSC &\ 1\\
0.89998 &\ 19.86 &\ 0.01 &\ D1.5m+DFOSC &\ 1\\
0.9037  &\ 19.88 &\ 0.01 &\ D1.5m+DFOSC &\ 1\\
0.90743 &\ 19.80 &\ 0.01 &\ D1.5m+DFOSC &\ 1\\
0.91146 &\ 19.92 &\ 0.01 &\ D1.5m+DFOSC &\ 1\\
1.82138 &\ 21.45 &\ 0.06 &\ D1.5m+DFOSC &\ 1\\
\hline
0.0179  &\ 20.29 &\   0.34 &\ SSO &\ 2\\
0.0221  &\ 19.95 &\   0.22 &\ SSO &\ 2\\
0.0262  &\ 19.95 &\   0.22 &\ SSO &\ 2\\
0.0303  &\ 19.95 &\   0.22 &\ SSO &\ 2\\
0.0713  &\ 19.10 &\   0.10 &\ SSO &\ 2\\
0.1188  &\ 19.20 &\   0.11 &\ SSO &\ 2\\
0.1608  &\ 19.10 &\   0.10 &\ SSO &\ 2\\
0.1968  &\ 18.99 &\   0.10 &\ SSO &\ 2\\
0.2427  &\ 18.79 &\   0.10 &\ SSO &\ 2\\
\hline
0.28938 &\ 18.92 &\   0.17 &\ Watcher &\ 3\\
0.30887 &\ 18.58 &\   0.11 &\ Watcher &\ 3\\
0.36713 &\ 18.97 &\   0.10 &\ Watcher &\ 3\\
0.38661 &\ 18.96 &\   0.11 &\ Watcher &\ 3\\
0.40613 &\ 19.04 &\   0.11 &\ Watcher &\ 3\\
0.33131 &\ 19.33 &\   0.30 &\ Watcher &\ 3\\
1.33851 &\ 20.69 &\   0.21 &\ Watcher &\ 3\\
1.47729 &\ 20.75 &\   0.26 &\ Watcher &\ 3\\
2.33615 &\ 21.27 &\   0.86 &\ Watcher &\ 3\\
\hline
0.59715 &\ 19.42 &\   0.03 &\ VLT+FORS2 &\ 4\\
0.59885 &\ 19.43 &\   0.02 &\ VLT+FORS2 &\ 4\\
0.59989 &\ 19.42 &\   0.02 &\ VLT+FORS2 &\ 4\\
0.60094 &\ 19.45 &\   0.02 &\ VLT+FORS2 &\ 4\\
0.602   &\ 19.45 &\   0.02 &\ VLT+FORS2 &\ 4\\
0.60313 &\ 19.43 &\   0.02 &\ VLT+FORS2 &\ 4\\
0.60418 &\ 19.46 &\   0.02 &\ VLT+FORS2 &\ 4\\
0.60524 &\ 19.45 &\   0.03 &\ VLT+FORS2 &\ 4\\
0.6063  &\ 19.45 &\   0.02 &\ VLT+FORS2 &\ 4\\
0.8701  &\ 19.88 &\  0.03  &\ VLT+FORS2 &\ 4\\
0.89996 &\ 19.92 &\  0.02  &\ VLT+FORS2 &\ 4\\
1.72583 &\ 21.07 &\  0.02  &\ VLT+FORS1 &\ 4\\
1.86974 &\ 21.25 &\  0.02  &\ VLT+FORS1 &\ 4\\
2.84199 &\ 22.47 &\  0.06  &\ VLT+FORS1 &\ 4\\
3.86899 &\ 23.04 &\  0.09  &\ VLT+FORS1 &\ 4\\
4.84365 &\ 23.58 &\  0.19  &\ VLT+FORS1 &\ 4\\
6.74083 &\ 24.32 &\  0.30  &\ VLT+FORS1 &\ 4\\
10.81441&\ 25.40 &\  0.77  &\ VLT+FORS1 &\ 4\\
14.77259&\ 25.78 &\  1.08  &\ VLT+FORS1 &\ 4\\
%19.67818&\ 25.78 &\  1.22891\\
%23.80494&\ 25.40 &\  1.13322\\
\hline
\enddata
{\label{0614Optical}}

\tablenotetext{a}{Time since BAT trigger.}

\tablenotetext{b}{The Vega magnitude is after correction for the Galactic extinction of
$E(B-V)=0.057$ mag, and subtraction of the host contribution, $R_{\rm host}=22.46\pm0.04$.}

\tablerefs{(1)\citealt{Fynb06}; (2) \citealt{Schmidt06}; (3) This work; (4)
\citealt{DellaValle06}. }

\end{deluxetable}
\clearpage

\begin{deluxetable}{lllllll}
\tablecolumns{7}
\tablewidth{0pt}
\tablecaption{Results (main parameters) of fits to the three spectral energy distributions of
GRB\,060614 at epochs 0.187, 0.798, and 1.905 days. For the broken power law
models we fit both with the power law slopes free, and for the case of a cooling break
where $\Gamma_1 = \Gamma_2 - 0.5$ (where $\Gamma = \beta + 1$). Galactic absorption,
$N_{\rm H,Gal}$, is fixed at 1.87$\times$10$^{20}$ cm$^{-2}$ \citep{Kalberla} and
Galactic extinction, $E(B-V)_{\rm Gal}$, is fixed at 0.057 mag (\citealt{Schlegel98}). Solar
metallicity is assumed in the X-ray absorption model and the extinction is modeled with
an SMC extinction law \citep{Pei}. All errors are quoted at the 90\% confidence level.}
\tablehead{
\colhead{Model} & \colhead{$E(B-V)$} & \colhead{$N_{\rm H}$} & \colhead{$\Gamma_1$} &
\colhead{$E_{\rm bk}$} & \colhead{$\Gamma_2$} & \colhead{$\chi^2$/dof} \\
\colhead{} & \colhead{mag} & \colhead{(10$^{22}$ cm$^{-2}$)}& \colhead{} & \colhead{keV} &
\colhead{} & \colhead{}}

\startdata

Epoch 1 & & & & & & \\

PL+SMC &  $<$0.02 & $<$0.02 & 1.75$\pm$0.02 & - & - & 60/68 \\
BKNPL+SMC & $<$0.2 & 0.03$^{+0.03}_{-0.02}$ & 0.9$\pm$0.4 &
0.005$^{+0.010}_{-0.002}$ & 1.9$\pm$0.1 & 49/66 \\
BKNPL+SMC & $<$0.04 & 0.03$^{+0.02}_{-0.01}$ & $\Gamma_2$-0.5 & 0.012$^{+0.001}_{-0}$ &
1.86$^{+0}_{-0.02}$ & 50/67\\ \hline

Epoch 2 & & & & & & \\
PL+SMC & $<$0.04 & $<$0.03 & 1.78$^{+0.02}_{-0.01}$ & - & - & 42/36 \\
BKNPL+SMC & 0.3$^{+0.1}_{-0.2}$ & 0.10$^{+0.04}_{-0.05}$ & $<$1.3 &
0.005$^{+0.025}_{-0.001}$ & 2.2$\pm$0.2 & 27/34 \\
BKNPL+SMC & $<$0.08 & 0.09$^{+0.03}_{-0.05}$ & $\Gamma_2$-0.5 & 0.2$^{+0.7}_{-0.19}$ &
2.1$^{+0.2}_{-0.1}$ & 31/35 \\ \hline

Epoch 3 & & & & & & \\
PL+SMC & 0.09$^{+0.06}_{-0.05}$ & 0.06$^{+0.03}_{-0.02}$ & 1.81$\pm$0.04 & - &
- & 26/35 \\
BKNPL+SMC & 0.2$\pm$0.1 & 0.08$^{+0.04}_{-0.03}$ & $>$0.1 & unbounded &
1.9$\pm$0.1 & 24/33 \\
BKNPL+SMC & 0.13$^{+0.06}_{-0.07}$ & 0.07$\pm$0.03 & $\Gamma_2$-0.5 & $<$0.008 &
1.86$^{+0.07}_{-0.08}$ & 24/34 \\ \hline
\enddata
\label{sedresults}
\end{deluxetable}

\clearpage
\begin{deluxetable}{cccccccc}
\tablecolumns{7} \tablewidth{0pt} \tablecaption{Comparison of power-law spectral
evolution for GRB\,060614 and GRB\,080503.} \tablehead{
\colhead{GRB} & \colhead{Time Interval} & \colhead{Photon Index} & \colhead{Index Error$^a$} & \colhead{$\chi^2/{\rm dof}$} & \colhead{Bandpass}& \colhead{Ref}\\
\colhead{}    & \colhead{s}           & \colhead{}             & \colhead{} &
\colhead{} & \colhead{keV} &\colhead{}}

\startdata
GRB\,060614 & -2.83-5.62 &1.63  &0.07  &$48.2/56$ & 15-150 &1 \\
--          & 5.62-97.0  &2.21  &0.04  &$40.9/56$ & 15-150 &1 \\
--          & 97.0-176.5 &2.37  &0.13  &$42.6/56$ & 15-150 &1 \\
\hline
GRB\,080503 & 0.2-0.6 &1.74  &0.28  &$78.1/58$ & 15-150 &2 \\
--          & 10-200  &1.93  &0.14  &$38.3/58$ & 15-150 &2 \\
--          & 0.0-0.7 &1.59  &0.28  &$69/59$   & 15-150 &3 \\
--          & 10-170  &1.91  &0.12  &$52/59$   & 15-150 &3 \\
--          & 81-282  &1.27  &0.03  &--        & 0.3-10 &3 \\
--          & 81-280  &1.33  &0.05  &$696.3/714$&0.3-10 &4 \\
--          & 83-107  &1.00  &0.13  &--        & 0.3-10 &5 \\
--          & 107-128 &1.11  &0.13  &--        & 0.3-10 &5 \\
--          & 128-150 &1.42  &0.14  &--        & 0.3-10 &5 \\
--          & 150-185 &1.66  &0.16  &--        & 0.3-10 &5 \\
--          & 185-256 &1.77  &0.16  &--        & 0.3-10 &5 \\
\hline
\enddata
{\label{080503}} \tablenotetext{a}{Errors are quoted at the $90\%$ confidence level.}
\tablerefs{ (1) \citealt{Mangano07}; (2) This work; (3) \citealt{Mao08}; (4) The UK Swift
Science Data Centre; (5) \citealt{Perley08}.}
\end{deluxetable}

\end{document}